\documentclass[apjl]{emulateapj}
\usepackage{apjfonts}

\slugcomment{}
\shorttitle{Ejection in T Tauri South}
\shortauthors{Loinard et al.}

\newcommand{\dechms}[4]{$#1^{\rm h}#2^{\rm m}#3\mbox{$^{\rm s}\mskip-7.6mu.\,$}#4$}

\newcommand{\decdms}[4]{$#1^{\circ}#2'#3\mbox{$''\mskip-7.6mu.\,$}#4$}

\newcommand{\Lsun}{L$_{\odot}$}
\newcommand{\Msun}{M$_{\odot}$}

\newcommand{\msec}[2]{$#1\mbox{$''\mskip-7.6mu.\,$}#2$}
\newcommand{\mdeg}[2]{$#1\mbox{$^{\circ}\mskip-7.6mu.\,$}#2$}

\slugcomment{}

\begin{document}

\title{Ejection of a Low Mass Star in a Young Stellar System in Taurus}
\author{Laurent Loinard\footnotemark[1], Luis F.\ Rodr\'{\i}guez\footnotemark[1] and Monica I.\ Rodr\'{\i}guez\footnotemark[1]}
\footnotetext[1]{Instituto de Astronom\'{\i}a, Universidad Nacional Aut\'onoma 
de M\'e\ xico, Apartado Postal 72--3 (Xangari), 58089 Morelia, Michoac\'an, 
M\'exico. l.loinard@astrosmo.unam.mx; l.rodriguez@astrosmo.unam.mx; 
m.rodriguez@astrosmo.unam.mx}

\begin{abstract}
We present the analysis of high angular resolution VLA radio observations, 
made at eleven epochs over the last 20 years, of the multiple system T Tauri. 
One of the sources (Sb) in the system has moved at moderate speed (5-10 km 
s$^{-1}$), on an apparently elliptical orbit during the first 15 years of 
observations, but after a close ($<$ 2 AU) encounter with the source Sa, it 
appears to have accelerated westward to about 20 km s$^{-1}$ in the last few 
years. Such a dramatic orbital change most probably indicates that Sb 
has just suffered an ejection -- which would be the first such event ever 
detected. Whether Sb will ultimately stay on a highly elliptical bound orbit, 
or whether it will leave the system altogether will be known with about five 
more years of observations.
\end{abstract}


\section{Introduction}

If they contain three or more members, the multiple stellar systems in which 
many stars are known to form (Abt \& Levy 1976; Poveda, Allen, \& Parrao 1982) 
can evolve chaotically (Valtonen \& Mikkola 1991). The least massive members 
of the system can then suffer an ejection (whereby they are launched on a 
highly elliptical -- but still bound -- orbit), or even an escape (a situation 
when the stars will leave the system altogether). Such violent events can have 
major implications for the subsequent stellar evolution. They can limit the 
accretion process that makes young stars grow and favor the production of 
brown dwarfs (Reipurth 2001; Reipurth \& Clarke 2001). This possibility, 
supported by detailed hydrodynamical simulations (Bate et al.\ 2002), could 
explain the flattening of the low-mass end of the initial mass function with 
respect to Salpeter's original power law.

\begin{figure*}
\centerline{\includegraphics[width=0.22\textwidth,angle=270]{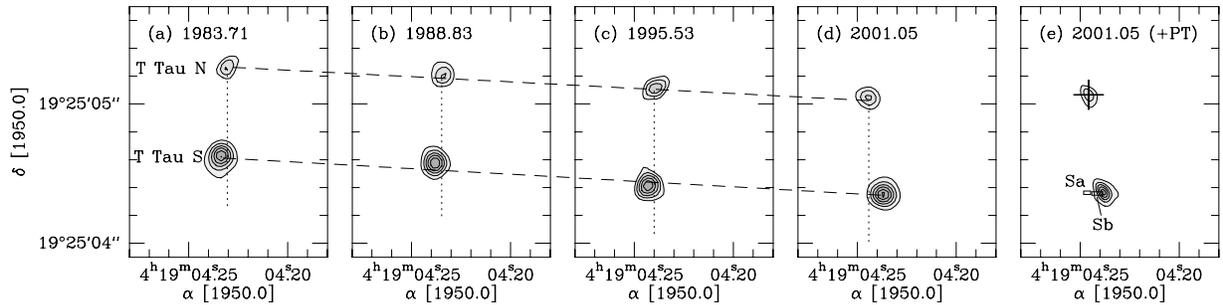}}
\caption{(a)--(d): Representative VLA 2 cm images of T Tauri at four of the 
epochs. The approximately horizontal dashed line has the slope given by the 
proper motion of the north component. The vertical dotted line is drawn at 
constant right ascension from the peak position of the north component. Note 
the proper motion of the system, as well as the rapid motion of the south 
component to the west between the last two epochs. (e): Comparison of the 
radio and infrared images for late 2000/early 2001. The two rectangles at the 
bottom of the panel indicate the error box on the position of Sa and Sb at 
the end of 2000, deduced from recent infrared images (Duch\^ene et al. 2002). 
The errors include the uncertainty on the pixel size in the reconstructed 
images ($\sim$ 0.6\%), and the uncertainty on the absolute orientation of the 
North ($\sim$ 1$^{\circ}$). The 2 cm radio image is that produced when Pie 
Town is included.}
\end{figure*}

Here we present the first direct evidence for this mechanism using multi-epoch
high angular resolution radio observations of the multiple system T Tauri.
T Tauri has long been known to be a binary system (Dyck, Simon, \& Zuckerman
1982). The north component, T Tau N, is considered the prototype for the whole 
class of low-mass pre-main sequence stars, and is indeed one of the best 
studied young stars in the sky. The southern infrared companion (T Tau S), 
located about 0.7 arcseconds (100 AU) to the south, is more obscured and not 
evident in the optical. Recent infrared images of T Tau S obtained by speckle 
holography showed that, in turn, it is composed of two stars, T Tau Sa and T 
Tau Sb, separated at the time of the first infrared observations (1997 -- 
Koresko 2000) by a mere 50 mas (7 AU). Subsequent observations (K\"ohler, 
Kasper, \& Herbst 2000; Duch\^ene, Ghez, \& McCabe 2002) showed that the 
separation between Sa and Sb is increasing, and that Sb is currently moving 
westward at the fairly large projected velocity of 20 km s$^{-1}$.

\section{Observations}

As part of an observational program devised to search for orbital motions in 
embedded young multiple systems, we reanalyzed archival observations of T 
Tauri taken with the Very Large Array of the US National Radio Astronomy 
Observatory (NRAO)\footnote{NRAO is a facility of the National Science 
Foundation operated under cooperative agreement by Associated Universtities, 
Inc.}. The eleven observations used here correspond to all the existing VLA 2 
cm observations obtained in the most extended configuration of the array (A) 
with an integration time larger than 1 hour; supplemented by a deep 3.5 cm 
integration obtained in 1998 also in the A configuration. The latter was 
included to have a continuous coverage of the source approximately every three 
years in the last two decades. The last 2 cm observation was obtained while 
Pie Town, an additional antenna which can be connected to the VLA to obtain 
higher resolution, was also in use. This dataset was reduced twice: once with, 
and once without Pie Town. 

The data were always recorded in both circular polarizations, with an 
effective bandwidth of 100 MHz. The sensitivity of the observations varies 
significantly from epoch to epoch but it is sufficient for the two components 
of the system (T Tau N and T Tau S) to be detected in all epochs but one; in 
the 1982 data, T Tau N is not detected. All observations used as phase 
calibrator the source 0400+258, whose absolute position has been refined over 
the years; the absolute registration used here corresponds to the oldest 
position for 0400+258: $\alpha(1950.0)$ = \dechms{04}{00}{03}{589}; 
$\delta(1950.0)$ = \decdms{25}{51}{46}{50}. Once calibrated, the 
interferometer visibilities were imaged using the IMAGR algorithm. To obtain 
relatively high angular resolutions, without losing much sensitivity, we used 
a weight intermediate between uniform and natural for all 2 cm observations; 
the resulting angular resolution is of the order of \msec{0}{13} at all epochs.
The 3.5 cm image was produced with uniform weighting to obtain the highest 
angular resolution possible ($\sim$ \msec{0}{20}). Finally, the synthesized 
beam for the 2001 2 cm image which includes Pie Town is \msec{0}{11} $\times$ 
\msec{0}{06} (P.A.\ 34$^{\circ}$).

\section{Results and discussion}

T Tauri is a double radio source (Schwartz, Simon, \& Campbell 1986, see also 
our Fig.\ 1), separated by about 0.7 arcsec in the north-south direction. A 
third radio component (T Tau R) which was reported from longer wavelength 
images (Ray et al.\ 1997), is never seen in the present data. The emission 
observed at 2 and 3.5 cm is probably largely dominated by free-free emission 
from ionized winds (Phillips, Lonsdale, \& Feigelson 1993), although 
gyrosynchrotron emission from a magnetically active photosphere 
might also contribute (Skinner \& Brown 1994; Johnston et al. 2003).

Although faint extended emission, presumably tracing episodic ejection events, 
is sometimes seen around T Tau S, the cores of both T Tau N and T Tau S remain 
essentially Gaussian at our angular resolution (see Fig.\ 1 for images at 
four of the epochs). This implies that extended components of the outflow 
powered by T Tau S (van Langevelde et al.\ 1994; Solf \& B\"ohm 1999) never 
dominate the short wavelength radio emission and are unlikely to affect 
significantly our astrometry. Since T Tau S appears as a single source, it is 
{\it a priori} unclear whether both components (T Tau Sa and Sb) known to 
exist there contribute to the total observed emission, or whether one of the 
two components strongly dominates the flux density. Fortunately, the last 
radio observation considered here is almost simultaneous with a recent 
infrared observation (Duch\^ene et al.\ 2002). As Fig.\ 1e shows, at that 
epoch, the radio source T Tau S is coincident within the errors with the 
infrared source T Tau Sb. 

A quick inspection of Fig.\ 1 immediately shows significant large-scale 
motions, with both components of the system moving towards the south--east. 
To characterize these motions, we determined the positions of the two sources 
at each epoch using Gaussian fits, and corrected these positions for the 
effect of the trigonometric parallax: a maximum of 7 mas, almost entirely in 
right ascension. The resulting motion of T Tau N (Fig.\ 2) appears to be 
remarkably linear, with translational proper motions of $\mu_\alpha$cos 
$\delta$ = 12.2 $\pm$ 0.6 mas yr$^{-1}$; $\mu_\delta$ = --12.7 $\pm$ 0.6 mas 
yr$^{-1}$, very similar to the values given in the Hipparcos catalog 
(Perryman et al.\ 1997) and by Jones \& Herbig (1979). Following previous
works (Loinard 2002; Loinard et al.\ 2002; Rodr\'{\i}guez et al.\ 2003; Curiel 
et al.\ 2003), we interpret these linear motions as due to the overall motion 
with respect to the Sun of the region where T Tauri is located. 

The motion of T Tau S is more complex: it follows the same trend as T Tau N 
until about 1992, but around 1995, it starts to move westwards (Fig.\ 2c). 
Because it is so close to T Tau Sa, T Tau Sb is most likely to be in 
gravitational interaction with it, so it is important to determine the motion 
of T Tau Sb relative to T Tau Sa. Since T Tau Sa is not actually seen in the
radio images, this involves two steps: first correcting the absolute motion 
of T Tau Sb for the large scale, nearly linear, motion of T Tau N; and then 
correcting for the motion of T Tau Sa relative to T Tau N. The former step is 
easily done by subtracting, for each epoch, the absolute position of T Tau N 
(Fig.\ 2a,b) to the absolute position of T Tau Sb. The latter step can be done 
fairly accurately because the relative motion of T Tau N and T Tau Sa is 
relatively well constrained thanks to multi-epoch near-infrared observations. 
Using all the available data (Ghez et al.\ 1995; Roddier et al.\ 2000; 
Duch\^ene et al.\ 2002), we estimate that the orbital motion of T Tau Sa 
around T Tau N results in rates of change of +\mdeg{0}{3} yr$^{-1}$ for the 
position angle, and of --3.4 mas yr$^{-1}$ for the separation. This is a 
somewhat smaller rate of change for the angular position than reported by 
Roddier et al.\ (2000). The difference comes from our taking into account the 
latest observation by Duch\^ene et al.\ (2002). At the epoch of that 
observation (late 2000), the position angle between N and Sa was only 
$179\rlap.^\circ7$, significantly less than the extrapolation from the data 
of Roddier et al.

Relative to Sa, the radio source T Tau S appears to first follow a roughly
elliptical path, then approach T Tau Sa around 1995--1998, and then rapidly
move away from it between 1998 to 2001 (Fig.\ 3). During the last 5.5 years of 
observation, it has moved by about 150 mas; at a distance of 140 pc, this 
proper motion corresponds to a projected velocity of $\sim$ 20 km s$^{-1}$, 
very similar to the projected velocity of T Tau Sb reported by Duch\^ene et 
al.\ (2002). This further confirms the identity between the radio source 
T Tau S and the infrared source T Tau Sb. It also implies that the motion 
seen in the radio does not trace a shock phenomenon (a Herbig-Haro-like 
nebulosity) but truly a stellar motion since T Tau Sb is known from infrared 
spectroscopy (Duch\^ene et al.\ 2002) to be a pre-main sequence M star.

\begin{figure*}
\centerline{\includegraphics[width=0.48\textwidth,angle=-90]{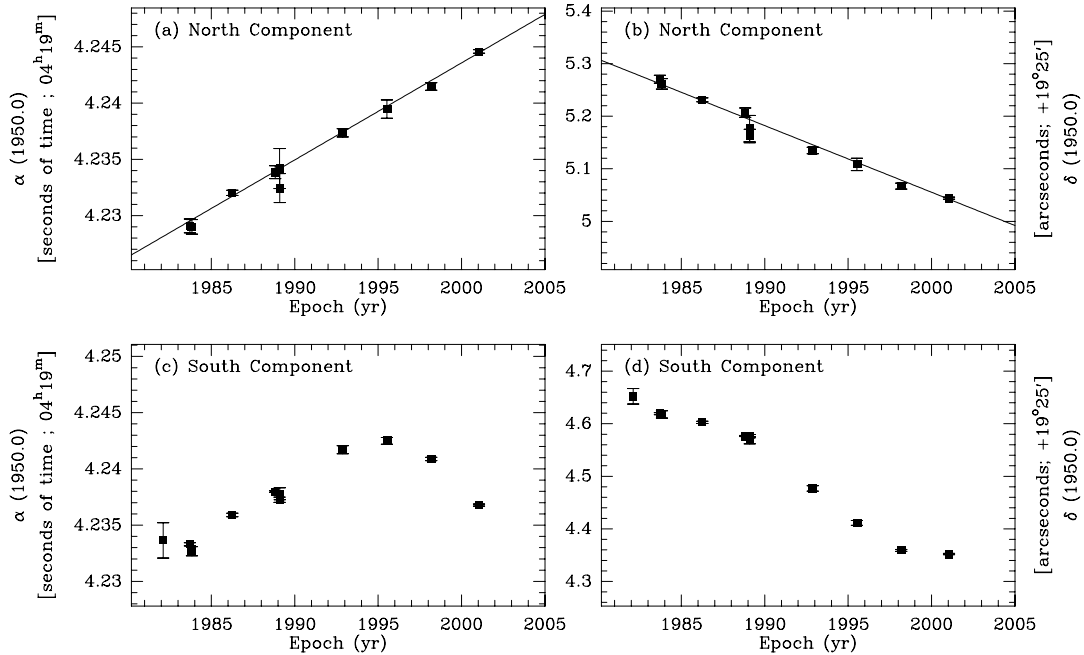}}
\caption{Proper motions of the T Tauri system in right ascension (left) and 
declination (right), for T Tau N (top) and T Tau S (bottom). The proper motion 
of T Tau N appears to be remarkably linear. In constrast, the proper motion 
of T Tau S shows a more complex behavior that we interpret as due to its 
ejection from the system. The errors affecting the measurements were 
computed as 0.6 times the ratio between the angular resolution and the 
signal-to-noise ratio in each image (Condon 1997).}
\end{figure*}

From 1983 to 1995, the motion of T Tau Sb relative to T Tau Sa appears to 
follow Kepler's second law: roughly equal areas are swept during equal times. 
This suggests that the first part of the motion does reflect a bound,
elliptical orbit. Under this assumption, one can estimate the mass of the T 
Tau S system using Kepler's third law. Given the projected semi-major axis 
(a $\sim$ 9 AU) of the orbit, and the apparent period (P $\sim$ 20 yr) a lower 
limit of 2 \Msun\ is required. Since Sb is a pre-main sequence star of type
M1 (Duch\^ene et al.\ 2002) its mass is about 0.5 \Msun. Hence, Sa must have 
a mass larger than about 1.5 \Msun. The second part of the path, on the other 
hand, follows a significantly different trend. While the distance between Sa 
and Sb changes by about the same amount from 1998 to 2001, as from 1989 to 
1995, the more recent part has been covered at twice the velocity of the first 
part (20 km s$^{-1}$ instead of 10 km s$^{-1}$ -- Fig.\ 3). Interestingly, 
the total mass required to keep the motion bound given the separation and 
velocity in late 2000 is at least 5 \Msun. This mass is quite larger than 
would be expected for such a low-luminosity young object. T Tau S has a 
bolometric luminosity of 11 \Lsun\ (Koresko, Herbst, \& Leinert 1997), that 
can be provided by a single pre-main sequence star with a mass of 1 to 2 
\Msun\ (Kenyon \& Hartmann 1995). A 5 \Msun\ pre-main sequence star is
expected to have a bolometric luminosity of several hundred Solar luminosities,
an order of magnitude higher than the measured luminosity of T Tau S.

Fig.\ 3 also shows that the radio source was close to the nominal position 
of Sa only between 1995 and 1998. At all other epochs, the separation between 
Sa and the radio source was large enough that the radio source could not have 
been Sa. This implies that the radio source is very likely to have been Sb
during the entire time span of the VLA observations, and that we are not
witnessing a ``Christmas Tree'' effect where the radio source corresponds 
sometimes to Sa and sometimes to Sb. The existence of one or more additional
radio sources which would have switched on and off during the observations 
appears as a particularly unlikely and {\it ad hoc} possibility. The analysis 
of the polarization of the 2 cm radio emission of T Tau S (Johnston et al.\ 
2003) further confirms this point since T Tau S is found to have consistently 
exhibited of the order of 10\% of left circular polarization (LCP) over the 
entire time-span covered by the observations.

Between 1995 and 1998, Sb has passed within about 15 mas (2 AU)
of Sa (Fig.\ 3). Such close encounters between a single low-mass star and a 
double system can lead to the ejection of the low-mass object, with a 
velocity of the order of the escape velocity at that separation (Sterzik \& 
Durisen 1998; Anosova 1986). For a 1.5 \Msun\ system, the escape velocity at 
2 AU is 35 km s$^{-1}$, about twice the current projected velocity of Sb. The 
radio source T Tau S (or, equivalently, the infrared source T Tau Sb) is, 
therefore, likely to be currently moving westward as a result of its recent 
ejection from the system. In this scheme, T Tau Sa would be a close binary 
system with a separation smaller than about 2 AU, and with a total mass 
of about 1.5\ \Msun; such a system would have an orbital period of 
approximately 2 years. Two pre-main sequence stars, each with a mass of 
about 0.5 to 1 \Msun\ would provide the required mass, and would have a total 
luminosity compatible with that of T Tau S.

As mentioned earlier, the first part of the orbit of Sb around Sa (1983 to 
1995) can be fitted by an ellipse (dotted curve in Fig.\ 3). Any subsequent 
large departure from this projected ellipse has to be taken as a change in the 
orbit, so from 1998 to 2001 Sb must indeed have changed its original orbit. 
Could this change be the result of a slight modification of the orbital plane 
due, for example, to viscous effects or to gravitational interaction with 
gaseous structures in the surrounding inhomogeneous medium? For motions nearly 
along the line of sight, small angular changes in direction can indeed result 
in large projected effects. However, the motions considered here cannot be 
nearly along the line of sight: the observed motions from 1998 to 2001 already 
imply a mass that is quite large given the luminosity of T Tau S. To maintain 
the possibility of bound motion, we have to assume that the velocity vector 
from 1998 to 2001 is close to the plane of the sky. In this configuration, any
small modification of the orbital plane will result in small projected changes.
So, could we have observed a large change in the direction of the motion, 
without a significant change in the magnitude of the velocity vector? Although 
possible, it is a fairly unlikely event. Equally large accelerations are 
needed to change significantly the direction or the magnitude of a velocity 
vector. Why would all the acceleration go into changing only the direction of 
the velocity vector rather than changing both direction and magnitude? We 
favor the interpretation that there was a large change in the velocity vector 
(quite likely both in magnitude and in direction) and that Sb is in a very 
different, wider orbit leading to an ejection or even possibly an escape. 

A statistical objection could be made against this interpretation. The 
probability of detecting an ejection in a young multiple system appears, at 
first sight, to be quite small. Non-hierarchical triple systems are inherently 
unstable and will eject a member on timescales of the order of one hundred 
crossing times (Anosova 1986). A system like T Tau S has a crossing time of 
order 10 years, so one should have to monitor about a hundred such systems 
over the timescales used in this paper to have a probability of order one to 
see an ejection. Yet, only a handful of such systems has been monitored, so 
we seem to be witnessing an event with a probability of only a few percent. 
However, important selection effects could be at work. The systems that 
undergo ejection events must have had previously close encounters in which 
all three members met in a small volume of space. Such close encounters will 
strongly enhance disk accretion and thus increase luminosity and outflow 
activity in the participating young stars. This is precisely the type of 
``active'' stars that are studied more systematically. Indeed, some of the 
mysterious properties of infrared companions (such as T Tau S) could plausibly 
be related to such enhanced activity. We may then be selecting {\it a priori} 
systems which have undergone close encounters in the recent past, and which 
are more likely to suffer an ejection. 

\begin{figure}
\centerline{\includegraphics[width=0.43\textwidth,angle=0]{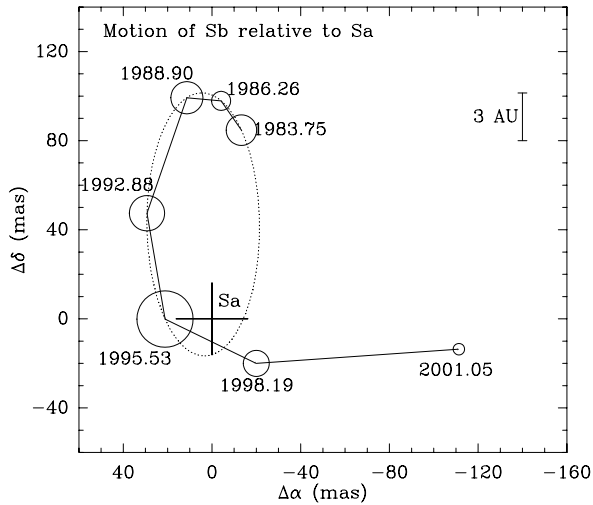}}
\caption{Orbit of T Tau Sb with respect to T Tau Sa. The numbers give the 
epoch of the observation and the circles the errors in the relative positions. 
For this figure we have made weighted averages of the observations made in 
near epochs (1983.71+1983.83=1983.75 and 1988.83+1989.09+1989.10=1988.90).
To give an idea of the sizes involved, a bar representing 3 AU is shown at 
the top right. The position of Sa (at (0,0)) is indicated by a cross, the size
of which represents the uncertainty on its position. The dotted ellipse is
a fit to the data from the first five epochs (1983.75 to 1995.53).}
\end{figure}

Whether the change of orbit of T Tau Sb described here will ultimately lead to
a full escape will be known if monitoring programs of T Tauri in the near 
infrared and the radio are performed in the coming 5 years. An argument 
favoring the escape scenario is related to the relatively large eccentricity 
of the true orbit of T Tau Sb, that following Hilditch (2001), we can estimate 
to be $e$ = 0.8 $\pm$ 0.2. For an elliptical orbit of eccentricity $e$ the 
velocity at periastron has to increase by $\Delta V/V~=~[2/(1+e)]^{1/2} - 1$ 
for the motion to become unbound. For $e \simeq 0.8$, a relatively small 
increase of 5\% in the velocity will change the orbit from elliptical to 
parabolic. However, given the short timescale of the ejection events, one 
should have {\it a priori} a higher chance of seeing the {\bf result} of the 
escape (i.e.\ a single star moving relatively fast compared to its environment)
than the escape itself. Yet, there is only one known T Tauri star which is 
moving fast compared to its surroundings (RW Aur -- Jones \& Herbig 1979), and 
a few indirect evidences for fast moving younger sources (Arce \& Goodman 
2002; Reipurth et al.\ 1999) or main sequence stars (Allen, Poveda, \& Worley 
1974; Hoogerwert, de Bruijne, \& de Zeeuw 2000). We speculate that this 
relative lack of detected ``fast-moving'' stars, however, might also be a 
consequence of observational biases. If the escape occurs, the pre-main 
sequence star will likely be left with very little of its original surrounding 
material, so the accretion and ejection activity will abruptly be halted. 
Within a very short time after the escape, the low-mass star is likely to 
reach a quiescent state, in which it would easily go unnoticed. Moreover, a 
star which would escape with a velocity only slightly above the escape 
velocity would have a small velocity once it reaches large distances from its 
original location (recall that a star escaping with exactly the escape speed 
will have zero velocity once at infinity). It would have truly large velocities
only for the short period of time just subsequent to the ejection. 

\acknowledgements
We acknowledge enriching discussions with Gaspard Duch\^ene and Arcadio Poveda.
Comments made by Bo Reipurth, and the referee, Alyssa Goodman, significantly 
improved the quality of this paper. We thank Barry Clark and the VLA data 
analysts for their help with the data retrieval process. We acknowledge 
support from CONACyT and DGAPA/UNAM.

\end{document}